\documentclass[useAMS,usenatbib]{mn2e}

\usepackage{graphicx}

\def\aj{AJ}%
\def\araa{ARA\&A}%
\def\apj{ApJ}%
\def\apjl{ApJ}%
\def\apjs{ApJS}%
\def\aap{A\&A}%
\def\aapr{A\&A~Rev.}%
\def\mnras{MNRAS}%
\newcommand{\avmass}{\langle M \rangle}
\newcommand{\mmin}{M_{\rm min}}
\newcommand{\mmax}{M_{\rm max}}

\newcommand{\msun}{\,{\rm M}_\odot}
\newcommand{\halfmass}{R_{\rm hm}}
\newcommand{\rvir}{R_{\rm vir}}
\newcommand{\mcl}{M_{\rm cl}}
\newcommand{\mcut}{M_{\rm cut}}
\newcommand{\trelax}{t_{\rm rlx}}

\title[How does the IMF affect star cluster evolution?]{How does a low-mass cut-off in the stellar IMF affect the evolution of young star clusters?}
\author[Kouwenhoven, Goodwin, de Grijs, Rose \& Kim]{
\parbox[t]{\textwidth}{\raggedright 
M.B.N. Kouwenhoven$^{1,2}$\thanks{E-mail:kouwenhoven@pku.edu.cn},
S.P. Goodwin$^{3}$, R. de Grijs$^{1,2}$, M. Rose$^{3,4}$, and~Sungsoo~S.~Kim$^{5}$}\\
\vspace{2pt}\\
   $^1$ Kavli Institute for Astronomy and Astrophysics, Peking
   University, Yi He Yuan Lu 5, Hai Dian District, Beijing 100871,
   China \\
   $^2$ Department of Astronomy, Peking
   University, Yi He Yuan Lu 5, Hai Dian District, Beijing 100871,
   China\\
   $^3$ Department of Physics \& Astronomy, The University of
   Sheffield, Hicks Building, Hounsfield Road, Sheffield S3 7RH, UK\\
   $^4$ Harvard-Smithsonian Center for Astrophysics, 60 Garden Street, Cambridge, MA 02138, USA\\
   $^5$ Department of Astronomy \& Space Science, Kyung Hee University,
   Yongin-shi, Kyunggi-do 446-701, Republic of Korea}

\begin{document}

\date{Accepted 2014 September 8. Received 2014 August 20; in original form 2014 July 22}

\pagerange{\pageref{firstpage}--\pageref{lastpage}} \pubyear{---}

\maketitle

\label{firstpage}


\begin{abstract}
We investigate how different stellar initial mass functions (IMFs) can affect the mass loss and
survival of star clusters.  We find that IMFs with radically 
different low-mass cut-offs (between 0.1 and $2\msun$) do not change
cluster destruction time-scales as much as might be expected.
Unsurprisingly, we find that clusters with more high-mass stars lose
relatively more mass through stellar evolution, but the response to this
mass loss is to expand and hence significantly slow their dynamical
evolution.  We also argue  that it is very difficult, if not
impossible, to have clusters with different IMFs that are initially
`the same', since the mass, radius and relaxation times depend on each
other and on the IMF in a complex way.  We conclude that changing the IMF to be biased
towards more massive stars does speed up mass loss and dissolution,
but that it is not as dramatic as might be thought.
\end{abstract}

\begin{keywords}
Stars: mass function -- stars: low-mass -- stars: kinematics and dynamics -- open clusters and associations: general
\end{keywords}


\section{Introduction} \label{section:introduction}

Star clusters are used as tracers of stellar populations and past star
formation in galaxies.  A key ingredient of `reverse engineering' an
observed population to its initial conditions is knowing how rapidly
clusters lose mass and are destroyed \citep[see, e.g.,][]{lamers2005, degrijs2007,
  chandar2010, karl2011, bastian2012, baumgardt2013}. 

If a star cluster survives the first few million years, then it will 
evolve as a result of
two-body relaxation, stellar evolution, interaction with the Galactic
tidal field, close encounters with molecular clouds and the effects
of disk and bulge shocking. All these effects contribute to mass loss
and dissolution  \citep[e.g.,][]{meylan1997,
  fukushige2000, heggiehut, lamers2006letter}. Numerous studies have
investigated the long-term evolution and final dissolution of various
types of star clusters under different environmental conditions
\citep[e.g.,][]{portegieszwart1998, baumgardt2003, lamers2005,
  gieles2008, degrijs2012, shin2013}.

One issue that has received relatively little attention recently is
the effect of different IMFs on the evolution of star clusters.  Part
of the reason for this lack of interest is the general feeling that
the IMF is universal and does not vary among star clusters
\citep[e.g.,][]{bastian2010}.  However, as we shall describe below, there
is possibly some evidence for variations, and variations in IMFs are
often claimed, so it is worth investigating how star cluster evolution will
change given different IMFs.

Previous studies have shown that the long-term survival of star
clusters depends on the properties of the low-mass section ($\la$ a
few~$\msun$) of their IMF. When a deficit of low stellar masses
exists, or when the slope of the IMF is too shallow (i.e., when the
stellar mass distribution is top-heavy), star clusters will likely
disperse within a billion years of their formation
\citep[e.g.,][]{chernoff1987, chernoff1990, goodwin1997, smith2001,
  mengel2002}. \cite{kim2006} compared
the evolution of clusters with different lower mass cut-offs for the
Arches cluster. They find that clusters with the same upper IMF but
with two different mass cut-offs ($0.1\msun$ and $1\msun$) do not give
significantly different luminosity profiles for the Arches cluster at the current age.

Theoretical and observational arguments have been proposed suggesting
that the IMF may depend on environment \citep[for a review, see][]{bastian2010}. For example, the upper mass limit of the IMF
of a star cluster may depend on the environment in which it forms
\citep[e.g.,][]{reddish1978,vanbeveren1982,weidnerkroupa2006},
although observational selection effects can complicate the derivation
of such a relationship \citep{parker2007,
  maschberger2008}. Extragalactic studies also suggest
that the IMF may be more top- or bottom-heavy in different
environments \citep[e.g.,][and numerous others]{brewer2012,
  spiniello2012,dutton2012,zaritsky2012, dabringhausen2012,
  ferreras2013, goudfrooij2013, geha2013, lasker2013, smith2013,
  bekki2013, weidner2013, barnabe2013}.

Several studies have claimed observational evidence for
a top-heavy IMF or a lower-mass cut-off in the IMF in young star
clusters. \cite{mccrady2003} suggest that MGG-11, a star cluster in the starburst galaxy M82, shows evidence of a top-heavy
IMF, with a lack of low-mass stars ($M<1\msun$). \cite{mccrady2005}
also discuss a possible lower mass limit in M82-F. They explain their
observations using a top-heavy IMF with a lower mass cut-off at approximately
$2\msun$. \cite{smith2001} claimed that M82-F has a lower-mass
cut-off at $2-3\msun$, but \cite{bastian2007} show that
this may be explained by differential extinction. 
Another example is NGC\,1705-1, where \cite{sternberg1998} finds 
that the IMF must be flat
or truncated below $M<1\msun$. \cite{mengel2008} examine 
young star clusters in NGC\,4038/4039 and find
that their results can be explained by a significant range in possible
IMF slopes or low-mass cut-offs.  \cite{greissl2010}, on the other
hand, finds no evidence for a low-mass cut-off. Finally,
\cite{stolte2005} find that the present-day mass function in the
Arches cluster near the Galactic Centre is truncated below $6-7\msun$,
although \cite{kim2006}
attribute this result to a bump in the IMF around $6-7\msun$. 

In summary, whilst there is no definitive evidence of variations in
the IMF with environment \citep[see][]{bastian2012}, there are many
claims, and environments, where the observations are unclear.
Therefore, it is worth studying the effect of IMF variations on the
evolution of star clusters, and even if the IMF is truly universal in
all environments, this is still an interesting theoretical
investigation. 

In this paper we carry out numerical simulations of moderately sized star
clusters, focusing on the first 200~Myr of their evolution.  This article is organised
as follows. In Section~\ref{section:method} we describe our method and
assumptions. In Section~\ref{section:results} we study how cluster
evolution depends on the properties of the IMF, by comparing the
evolution of star clusters with varying initial conditions. We discuss the implications of our findings in Section~\ref{section:discussion} and finally we summarise our conclusions 
in Section~\ref{section:conclusion}.


\section{Method} \label{section:method}

We simulate ensembles of moderate-mass star clusters (typically a few
thousand solar masses) with different IMFs.  An important point that we
will keep returning to is that it is impossible to create two clusters
with different IMFs that are actually `the same' - at least one of the parameters
mass, radius or relaxation time will differ between clusters with
different IMFs, often significantly.

\subsection{Initial conditions} \label{section:initialconditions}

We simulate clusters with typical masses of around $1500\msun$ (although
this varies from 27 to $17700\msun$ for reasons we will describe
below).  Clusters are evolved with and without stellar evolution to
discriminate the dynamical evolution from that driven by stellar mass
loss.  We vary the lower-mass cut-off in the IMF between 0.1 and
$2\msun$.  In order to compare `like-with-like' we run various ensembles
in which we keep any two of the cluster mass, half-mass radius,
half-mass relaxation time, and the upper end of the IMF, constant.

We use the publicly available \texttt{NBODY6} package
\citep{aarseth2003} for our simulations. Stellar evolution 
and binary evolution are
integrated following the recipes of \cite{eggleton1989,eggleton1990},
\cite{tout1997} and \cite{hurley2000}.  

Each cluster starts as a Plummer sphere in virial equilibrium \cite[following][]{aarseth1974}, and the most massive star
allowed in any cluster is $20\msun$.    The
fundamental upper mass limit of the IMF may be as high as $300\msun$
\citep{crowther2010}, but what we are effectively doing is
ignoring the first few million years of the life of the cluster and starting
with a population of clusters that have survived any initial gas
expulsion phase, and relaxed into a bound cluster.  Therefore, whilst
we start our clusters at a formal age of zero, really the starting
point for our simulations is an age of $5-10$~Myr and any stars $>20\msun$ will have evolved.  This avoids
complications from what are the true initial conditions from star
formation \citep[such as initial substructure; see][]{allison2010}.  Also
note that $20\msun$ is the
maximum mass one would expect in our canonical $\mcl \approx
1500\msun$ cluster, either by random sampling \citep{parker2007}
or from a cluster mass-maximum stellar mass relationship \citep{weidnerkroupa2006}. It should be noted that, observationally, it is impossible to tell the difference between these two scenarios \citep{cervino2013}. 
We do not include primordial binaries, nor do we
consider primordial mass segregation. 

We define a canonical reference cluster with a mass of $\mcl =
1500\msun$, and a virial radius 
$\rvir=1$~pc, with a corresponding initial
projected half-mass radius $\halfmass = 0.59\rvir$ and an intrinsic
half-mass radius of $0.77\rvir$ \citep[see, e.g.,][]{heggiehut}. These
are typical sizes of young open clusters, although the observed spread
in radii is large \citep[e.g.,][]{lada2003, schilbach2006, portegieszwart2010}.

We sample a stellar mass distribution $f(M)$ in the mass range
$\mcut \leq M \leq \mmax$, where $\mcut$ is a varying low-mass cut-off
in the IMF.  We sample
$\mcut$ with values each separated by $\sqrt{2}$, i.e., equally in
logarithmic space: $\mcut \approx 0.10$, 0.14, 0.20, 0.32, 0.50, 0.71,
1.00, 1.41, and $2.00\msun$. The minimum value of $\mcut$ is near the hydrogen-burning limit, and the range of $\mcut$ roughly brackets the values claimed in observational studies. The properties of each of these models
are listed in Table~\ref{table:models}. 

We consider both a full Salpeter IMF
\citep[e.g.,][]{salpeter1955, oey2011} and the \cite{kroupa2001}
IMF. The Salpeter IMF is a power-law $f(M) \propto
M^{\alpha}$ with $\alpha=-2.35$. Subsequently, we adopt the more
realistic \cite{kroupa2001} IMF, a three-part power-law mass
distribution, which has $\alpha=-2.3$ at the high-mass end. Although
the Salpeter IMF is unrealistic down to the hydrogen burning limit,
the effect of a low-mass cut-off is very prominent, and it can
therefore be used to illustrate the general behaviour of clusters with
a low-mass cut-off in the IMF. The \cite{kroupa2001} IMF is used to
determine how much a low-mass cut-off affects more realistic
clusters.  For a simple power-law IMF, $f(M) \propto M^\alpha$, the
average mass $\avmass$ for the Salpeter IMF (with $\alpha=-2.35$), $\mmin = \mcut$ and
$\mmax=20\msun$, is
\begin{equation} \label{eq:avmass_kroupa}
  \avmass_{\rm S} \approx 3.86 \left( \frac{ 0.35 -\mcut^{-0.35}}{
    0.0175-\mcut^{-1.35}} \right)\msun \,.
\end{equation}
For the \cite{kroupa2001} IMF the average mass can be calculated
numerically, and we find that the following expression is a good
approximation:
\begin{equation}
  \avmass_{\rm K} \approx 0.35 + 2.23\mcut + 0.05\mcut^2 \,.
\end{equation}

We include the external tidal field of the host galaxy, assuming that
the cluster is on a circular orbit in the Solar neighbourhood. 
The Jacobi radius $r_{\rm J}$ of a star cluster of mass $\mcl$ at a
Galactocentric distance $D_{\rm G}$ can, to first order, be approximated by
\begin{equation} \label{eq:jacobi}
  r_{\rm J} \approx D_{\rm G} \left( \frac{\mcl}{3M_{\rm G}} \right)^{1/3}    \approx
  6.65 \left( \frac{\mcl}{1000\msun} \right)^{1/3} \ {\rm pc}      \,,
\end{equation}
\citep{binneytremaine}, where we adopt $M_{\rm G}=5.8\times 10^{11}\msun$ as
the mass of a Milky Way-like galaxy and $D_{\rm G} \approx 8$~kpc for the
Galactocentric distance.  For star clusters of mass $\mcl \approx
1500\msun$ (see Table~\ref{table:models}) the Jacobi radius is roughly
$r_{\rm J} \approx 7.6$~pc. 

As the clusters evolve, stars gradually escape through ejection or through interaction with the Galactic tidal field. Previous work has shown that simple escape criteria such as the binding energy and/or a distance beyond the Jacobi radius are not sufficient, as many stars satisfying these criteria can still spend a significant amount of time near the cluster and interact with neighbouring stars, or even return to the star cluster \citep[e.g.,][]{terlevich1987, fukushige2000, ross1997}. Loosening the escape criteria is a safer approach, but this also has the risk of retaining escaping stars for too long, which is problematic when the process of cluster mass loss is studied. Previous work has indicated that adopting an escaper criterion of twice the Jacobi radius (Eq.~\ref{eq:jacobi}) is a practical compromise \citep[e.g.,][]{aarseth1973, aarseth2003, portegies2001}, and this is also the approach we adopt in our study. The consequence of this choice is that we may identify escaping stars slightly too late. For example, when a star formally escapes the star cluster at a distance $r$ from the cluster centre at a radial orbit with velocity $v$, then it will be identified as as an escaper at a time $\Delta t\approx (2r_{\rm J}-r)/v$ later. Our modelled star clusters typically have $r_{\rm J} \approx 7.6$~pc, and most stars escape with $1-10$~km\,s$^{-1}$, such that $\Delta t < 1.5-15$~Myr. Although the escape rate and cluster membership are correctly calculated over longer time-scales, caution should be taken when interpreting differences in star cluster membership over shorter time-scales.

The total integration time for each model is 200~Myr, which is substantially longer than the three time-scales that determine the global evolution of the star clusters studied: the stellar evolutionary time-scales, the crossing time, and the relaxation time (see Section~\ref{section:comparison}).
Depending on the number of member stars in a cluster, we run between
tens and thousands of realisations of each model (keeping $N$ multiplied by the number of realisations roughly constant at $1.5\times 10^5$) to reduce statistical
fluctuations, which is especially important in some cases with very small-$N$.


\subsection{Dynamics and comparisons between clusters} \label{section:comparison}

The fundamental process we are interested in is the evolution of the
cluster mass with time -- i.e. how fast a cluster loses mass, and
hence its lifetime. We expect two processes to be important in the evolution of our
clusters.

First, and most obviously, stellar evolutionary mass loss will be 
important.  Stellar evolution
will cause stars to lose a significant fraction of their mass at the
end point of their evolution.  The time-scale at which stellar evolution becomes
important roughly corresponds to 10, 20, 50, 100, and 200~Myr
for stars of mass 17.5, 11, 6.8, 4.8, and $3.7\msun$, respectively:
high-mass stars lose more mass, more rapidly than low-mass stars.

So, the greater the fraction of the initial mass of a cluster that
is in higher-mass stars, the more mass that cluster will lose, and
the faster it will evolve. 

The effect of stellar evolutionary mass loss is to cause the cluster
to become less massive (obviously), and also to expand.  Expansion leads to
two effects, one hastens destruction, the other slows it.  Expansion
causes the crossing time and the relaxation time to increase, so it
slows down dynamical evolution and aids survival.  But expansion due
to mass loss causes the cluster to fill more of a now smaller tidal
radius and eases the loss of stars and hastens destruction. As we shall
see, the balance between these effects is important.

In most cases we take a Salpeter IMF between $\mcut$ and $\mmax=20\msun$ as
our IMF.  For a Salpeter IMF with a low-mass cut-off at $\mcut=0.1\msun$ the
cluster will lose approximately 10 per cent of its mass in 100~Myr, and approximately 15 per cent
by 200~Myr through stellar evolution alone. For $\mcut=1\msun$ the percentages are 28 and 35 at 100 and 200~Myr, and for $\mcut=2\msun$, 42
and 55 at 100 and 200~Myr, respectively.
So we would expect to see the masses of clusters fall by {\em at
  least} this amount in 100 and 200~Myr.  Any further mass loss must
be due to dynamics.

The other important process in cluster evolution is dynamics: 
interactions redistribute energy between stars 
and causes the loss of (preferentially low-mass) stars.  This can
occur in a violent close encounter, or simply by small perturbations
(and the input of tidal energy) causing a star to reach the escape
velocity and pass beyond the tidal radius \citep[e.g.,][]{heggiehut}. In addition, scattering events can also result in high-velocity ejections of massive stars (and sometimes even binaries) that have sunk to the centre of the star cluster as a result of mass segregation \citep[see, e.g.,][]{gualandris2004}, although the vast majority of massive stars evolve before this occurs, and leave the star clusters as stellar remnants (see Section~\ref{section:results}).

Dynamical interactions are driven by encounters between stars/stellar 
systems and the fundamental time-scale for encounters is the crossing
time:
\begin{equation}
t_{\rm cr} = \frac{R}{\sigma},
\end{equation}
where $R$ is the size of the system, and $\sigma$ the velocity
dispersion.  The half-mass crossing time in a virialised system is 
\begin{equation} 
t_{\rm cr(half)} = \sqrt{ \frac{2}{\rm G} \frac{\halfmass^3}{\mcl} }
\end{equation}
where $\halfmass$ is the half-mass radius, $\mcl$ the total cluster mass,
and ${\rm G}$ the gravitational constant.

Although we do not include primordial binaries, dynamical binaries may form through three-body encounters. If the cluster contains a binary system, then that binary can act as an energy
sink: encounters remove energy from the binary, making it `harder'
whilst decreasing (making less negative) the potential energy of the
rest of the cluster.  Close encounters with the binary can also cause 
ejections.  In a star cluster with a single energetically
important binary (usually near its centre), the encounter rate with
this binary system scales with the crossing time.

Two-body encounters between single stars in the cluster will cause
both energy equipartition/mass segregation and evaporation.
The global dynamical evolution of star clusters occurs at the
time-scale of relaxation. The half-mass relaxation time $\trelax$ for
a star cluster with a \cite{plummer1911} distribution is
\begin{equation} \label{eq:relaxationtime}
  \trelax \sim \left( \frac{N}{ 8 \ln N} \right) t_{\rm cr(half)}
\end{equation}
\citep{heggiehut}, where $N$ is the number of stars in the cluster
\citep[see, e.g.,][]{binneytremaine, chernoff1990}.

Therefore, there are three time-scales that determine the evolution of star clusters: 
\begin{enumerate}
\item The stellar evolutionary time-scale (the time-scale on which we lose a significant
amount of mass through stellar evolution), which depends on $\avmass$;
\item The crossing time, which depends on $\halfmass$ and $\mcl$; and
\item The relaxation time, which depends on the crossing time
(i.e., $\halfmass$ and $\mcl$), and also on $N$ (which depends on $\mcl$
and $\avmass$).
\end{enumerate}
Stars also escape when they pass beyond the tidal boundary, which
also depends on $\mcl$ and therefore shrinks as stars evolve and escape over time.
All cluster parameters will evolve with time: $\mcl$ and $N$ will
always decrease (but not at the same rate) as stars 
evolve or are ejected, but $\halfmass$ and
$\avmass$ can increase, decrease or stay roughly the same.  Therefore,
dynamical time-scales can evolve in complex ways.

\subsubsection{Comparing clusters}

When studying the effect of varying IMFs on the evolution of star 
clusters, one would ideally like to only vary one parameter:
$\mcut$, and hence $\avmass$.  However, as we have seen, changing 
$\avmass$ changes $N$, which changes the relaxation time.  Keeping the
relaxation time constant then forces us to change other parameters,
and so on.

Therefore we run several different sets of simulations, for each of which we keep the
initial conditions of several parameters constant while varying
$\mcut$. The different sets of models,
which we refer to as models MR, TR, MT and UR, respectively, are as
follows:
\begin{itemize}
\item {\em Model~MR}: The initial total cluster mass and initial
  half-mass radius are fixed (Section~\ref{section:modelm}).
\item {\em Model~TR}: The initial half-mass relaxation time 
  and initial half-mass radius are fixed
  (Section~\ref{section:modelt}).
\item {\em Model~MT}: The initial total mass and initial
  half-mass relaxation time are fixed
  (Section~\ref{section:modelr}). 
\item {\em Model~UR}: The upper part of the IMF and the initial
  half-mass radius are fixed (Section~\ref{section:modelk}). 
\end{itemize}
The last model, UR, requires some further explanation.  In this model
the numbers/masses of stars with masses above $3\msun$ are
kept constant.  This is in order to represent clusters that would `look'
similar to an observer (for more distant clusters, only the most
massive stars can be observed).  Therefore, a hypothetical observer
looking at any cluster in model~UR would see a cluster with the same
half-light radius and the same higher-mass stellar content.  They
might not be able to observe that the low-mass cut-off of the IMF
varied among these clusters.

The {\em initial} properties of each of the models are shown in
Table~\ref{table:models}: the identifier of the simulations, the initial mass function, the
low-mass cut-off $\mcut$, the total mass $\mcl$, the average stellar
mass $\avmass$, the number of stars $N$, the half-mass relaxation time
$\trelax$ and the half-mass radius $\halfmass$.


\begin{table}
  \caption{Initial conditions of the models used in our analysis. Simulations
    of each model are carried out with and without stellar evolution. The first column lists the model ID (see Section~\ref{section:comparison}). The
    adopted shape of the IMF (S = Salpeter, K = Kroupa) for each
    model is listed in the second column. The remaining columns list initial values of the cut-off mass $\mcut$, the total cluster mass $\mcl$, the average stellar mass $\avmass$, the total number of stars $N$, the half-mass relaxation time $\trelax$ and finally the half-mass radius $\halfmass$.
    \label{table:models} }
  \begin{tabular}{ccc r rrrr}
    \hline
    ID & IMF & $\mcut$ & \multicolumn{1}{c}{$\mcl$}  & \multicolumn{1}{c}{$\avmass$} & \multicolumn{1}{c}{$N$} & \multicolumn{1}{c}{$\trelax$} & \multicolumn{1}{c}{$\halfmass$} \\
    & & \multicolumn{1}{c}{$\msun$} & \multicolumn{1}{c}{$\msun$} & \multicolumn{1}{c}{$\msun$}  &     & \multicolumn{1}{c}{Myr} & \multicolumn{1}{c}{pc}\\
    \hline
MR1 & S & 0.10 & 1500.0 & 0.326 & 4607 & 19.64 & 0.59 \\
MR2 & S & 0.14 & 1500.0 & 0.445 & 3367 & 14.91 & 0.59 \\
MR3 & S & 0.20 & 1500.0 & 0.619 & 2424 & 11.19 & 0.59 \\
MR4 & S & 0.32 & 1500.0 & 0.948 & 1583 & 7.73 & 0.59 \\
MR5 & S & 0.50 & 1500.0 & 1.408 & 1065 & 5.50 & 0.59 \\
MR6 & S & 0.71 & 1500.0 & 1.908 & 786 & 4.24 & 0.59 \\
MR7 & S & 1.00 & 1500.0 & 2.550 & 588 & 3.32 & 0.59 \\
MR8 & S & 1.41 & 1500.0 & 3.383 & 443 & 2.62 & 0.59 \\
MR9 & S & 2.00 & 1500.0 & 4.467 & 336 & 2.08 & 0.59 \\
\hline
TR1 & S & 0.10 & 26.7 & 0.326 &      82 & 5.00 & 0.59 \\
TR2 & S & 0.14 & 62.8 & 0.445 &     141 & 5.00 & 0.59 \\
TR3 & S & 0.20 & 148.5 & 0.619 &    240 & 5.00 & 0.59 \\
TR4 & S & 0.32 & 436.1 & 0.948 &    460 & 5.00 & 0.59 \\
TR5 & S & 0.50 & 1151.9 & 1.408 &   818 & 5.00 & 0.59 \\
TR6 & S & 0.71 & 2392.4 & 1.908 &  1254 & 5.00 & 0.59 \\
TR7 & S & 1.00 & 4766.1 & 2.550 &  1869 & 5.00 & 0.59 \\
TR8 & S & 1.41 & 9261.9 & 3.384 &  2737 & 5.00 & 0.59 \\
TR9 & S & 2.00 & 17700.8 & 4.469 & 3961 & 5.00 & 0.59 \\
\hline
MT1 & S & 0.10 & 1500.0 & 0.326 & 4627 & 5.00 & 0.24 \\
MT2 & S & 0.14 & 1500.0 & 0.445 & 3389 & 5.00 & 0.28 \\
MT3 & S & 0.20 & 1500.0 & 0.619 & 2437 & 5.00 & 0.34 \\
MT4 & S & 0.32 & 1500.0 & 0.948 & 1591 & 5.00 & 0.44 \\
MT5 & S & 0.50 & 1500.0 & 1.408 & 1071 & 5.00 & 0.55 \\
MT6 & S & 0.71 & 1500.0 & 1.909 & 791  & 5.00 & 0.66 \\
MT7 & S & 1.00 & 1500.0 & 2.550 & 591  & 5.00 & 0.77 \\
MT8 & S & 1.41 & 1500.0 & 3.383 & 446  & 5.00 & 0.91 \\
MT9 & S & 2.00 & 1500.0 & 4.466 & 338  & 5.00 & 1.06 \\  
\hline
UR1 & K & 0.10 & 4664.6 & 0.565 & 8263 & 18.67 & 0.59 \\
UR2 & K & 0.14 & 4485.5 & 0.665 & 6750 & 15.91 & 0.59 \\
UR3 & K & 0.20 & 4242.9 & 0.800 & 5304 & 13.22 & 0.59 \\
UR4 & K & 0.32 & 3816.3 & 1.054 & 3620 & 9.96 & 0.59 \\
UR5 & K & 0.50 & 3254.5 & 1.462 & 2226 & 7.05 & 0.59 \\
UR6 & K & 0.71 & 2768.9 & 1.972 & 1404 & 5.13 & 0.59 \\
UR7 & K & 1.00 & 2341.2 & 2.622 & 893 & 3.78 & 0.59 \\
UR8 & K & 1.41 & 1956.6 & 3.463 & 565 & 2.81 & 0.59 \\
UR9 & K & 2.00 & 1602.1 & 4.551 & 352 & 2.09 & 0.59 \\
  \hline
  \end{tabular}
\end{table}

\begin{table}
  \caption{Star cluster properties after $t=200$~Myr for the initial conditions. The first two columns list the model ID and the cut-off mass $\mcut$. The remaining columns show, both for models with and without stellar evolution, the fraction of the initial number of stars remaining, $N(t)/N(0)$, the fraction of the total initial mass remaining, $M(t)/M(0)$, and the final half-mass radius $\halfmass$ Note that the data for model~TR1 (with stellar evolution) are missing, since the majority of these clusters dissolve before $t=200$~Myr.
    \label{table:results} }
  \begin{tabular}{ll c ccc ccc}
    \hline
    \# & $\mcut$ &\multicolumn{3}{l}{With stellar evolution} &
    \multicolumn{3}{l}{Without stellar evolution} \\
     & $\msun$  & $\frac{N(t)}{N(0)}$ & $\frac{M(t)}{M(0)}$ & $\frac{\halfmass}{\rm pc}$ & $\frac{N(t)}{N(0)}$
         & $\frac{M(t)}{M(0)}$ & $\frac{\halfmass}{\rm pc}$ \\
    \hline
    MR1 & 0.10 & 0.92 & 0.80 & 2.38 & 0.88 & 0.84 & 3.06 \\
MR2 & 0.14 & 0.91 & 0.77 & 2.46 & 0.87 & 0.82 & 3.32 \\
MR3 & 0.20 & 0.89 & 0.74 & 2.61 & 0.85 & 0.80 & 3.52 \\
MR4 & 0.32 & 0.87 & 0.69 & 2.78 & 0.82 & 0.79 & 3.75 \\
MR5 & 0.50 & 0.83 & 0.63 & 3.01 & 0.79 & 0.77 & 3.99 \\
MR6 & 0.71 & 0.80 & 0.57 & 3.25 & 0.77 & 0.75 & 4.05 \\
MR7 & 1.00 & 0.74 & 0.50 & 3.70 & 0.75 & 0.74 & 4.22 \\
MR8 & 1.41 & 0.65 & 0.41 & 4.31 & 0.73 & 0.73 & 4.28 \\
MR9 & 2.00 & 0.51 & 0.28 & 5.29 & 0.72 & 0.72 & 4.37 \\
\hline
TR1 & 0.10 & --- & --- & --- & 0.47 & 0.75 & 0.18 \\
TR2 & 0.14 & 0.18 & 0.18 & 2.33 & 0.38 & 0.59 & 0.33 \\
TR3 & 0.20 & 0.39 & 0.36 & 2.89 & 0.36 & 0.48 & 1.38 \\
TR4 & 0.32 & 0.71 & 0.57 & 3.12 & 0.53 & 0.55 & 4.53 \\
TR5 & 0.50 & 0.81 & 0.61 & 3.10 & 0.76 & 0.74 & 4.21 \\
TR6 & 0.71 & 0.83 & 0.60 & 3.08 & 0.82 & 0.79 & 3.77 \\
TR7 & 1.00 & 0.83 & 0.55 & 3.21 & 0.85 & 0.83 & 3.31 \\
TR8 & 1.41 & 0.79 & 0.49 & 3.51 & 0.87 & 0.85 & 3.13 \\
TR9 & 2.00 & 0.70 & 0.38 & 4.28 & 0.88 & 0.87 & 2.88 \\
\hline
MT1 & 0.10 & 0.84 & 0.73 & 2.27 & 0.83 & 0.76 & 2.65 \\
MT2 & 0.14 & 0.84 & 0.72 & 2.38 & 0.82 & 0.75 & 2.96 \\
MT3 & 0.20 & 0.84 & 0.71 & 2.57 & 0.81 & 0.76 & 3.29 \\
MT4 & 0.32 & 0.84 & 0.67 & 2.78 & 0.80 & 0.76 & 3.67 \\
MT5 & 0.50 & 0.83 & 0.63 & 2.99 & 0.79 & 0.76 & 3.93 \\
MT6 & 0.71 & 0.80 & 0.58 & 3.34 & 0.78 & 0.76 & 4.09 \\
MT7 & 1.00 & 0.76 & 0.51 & 3.76 & 0.77 & 0.77 & 4.19 \\
MT8 & 1.41 & 0.68 & 0.42 & 4.47 & 0.77 & 0.77 & 4.26 \\
MT9 & 2.00 & 0.52 & 0.29 & 5.76 & 0.77 & 0.78 & 4.40 \\
\hline
UR1 & 0.10 & 0.95 & 0.84 & 1.90 & 0.94 & 0.90 & 2.21 \\
UR2 & 0.14 & 0.94 & 0.83 & 1.94 & 0.94 & 0.89 & 2.23 \\
UR3 & 0.20 & 0.94 & 0.82 & 2.02 & 0.93 & 0.89 & 2.37 \\
UR4 & 0.32 & 0.93 & 0.80 & 2.08 & 0.92 & 0.88 & 2.61 \\
UR5 & 0.50 & 0.90 & 0.75 & 2.32 & 0.90 & 0.86 & 2.76 \\
UR6 & 0.71 & 0.87 & 0.71 & 2.53 & 0.87 & 0.84 & 3.12 \\
UR7 & 1.00 & 0.83 & 0.63 & 2.85 & 0.84 & 0.81 & 3.35 \\
UR8 & 1.41 & 0.75 & 0.53 & 3.32 & 0.79 & 0.77 & 3.77 \\
UR9 & 2.00 & 0.60 & 0.38 & 4.15 & 0.74 & 0.74 & 4.00 \\
\hline

  \end{tabular}
\end{table}



\section{Results} \label{section:results}


A reasonable expectation is that clusters with a high $\mcut$ (i.e., a
lack of low-mass stars) will lose mass more rapidly and be destroyed
more rapidly that those with a low-$\mcut$.  

High-$\mcut$ clusters inevitably lose more mass through stellar
evolution than low-$\mcut$ clusters.  But the key question of interest is how this extra
(evolutionary) mass loss changes the rate at which dynamical mass loss
or tidal overflow occurs and so changes the rate at which the cluster
is destroyed.  In almost all cases we show that the extra evolutionary
mass loss does not have as significant an effect as one might expect.


\subsection{Identical $\mcl$ and $\halfmass$ (model~MR)} \label{section:modelm}

\begin{figure*}
  \centering
  \includegraphics[width=1\textwidth,height=!]{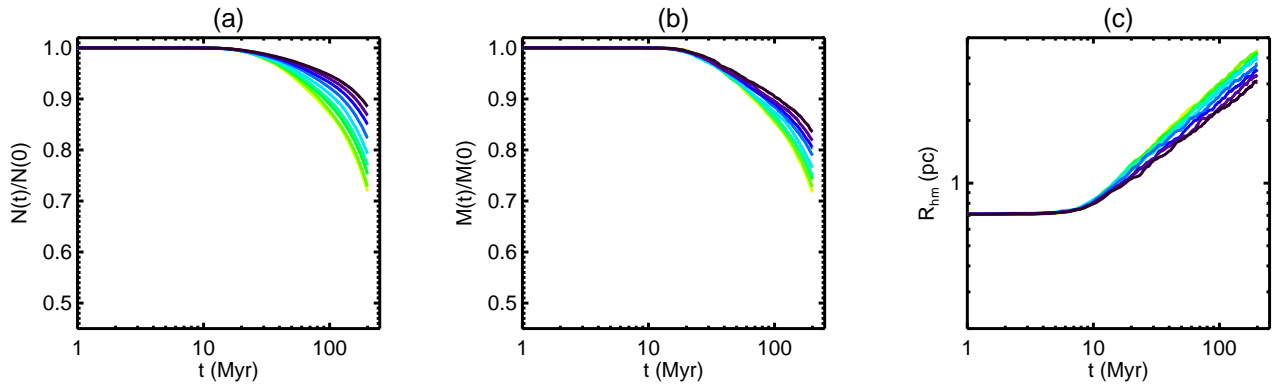}\\
  \caption{Evolution of star clusters with constant initial mass
    and half-mass radii (model~MR) without stellar evolution. (a) Fractional
    evolution of the number of stars, $N$. (b) 
    Fractional evolution of the cluster mass, $\mcl$. (c) 
    Evolution of the half-mass radii, $\halfmass$.  In each panel
    the darkest curves are for IMF low-mass cut-offs of $0.1\msun$,
    becoming lighter as the mass of the low-mass cut-off increases to $2\msun$.
    \label{fig:mn}}
\end{figure*}
\begin{figure*}
  \centering
  \includegraphics[width=1\textwidth,height=!]{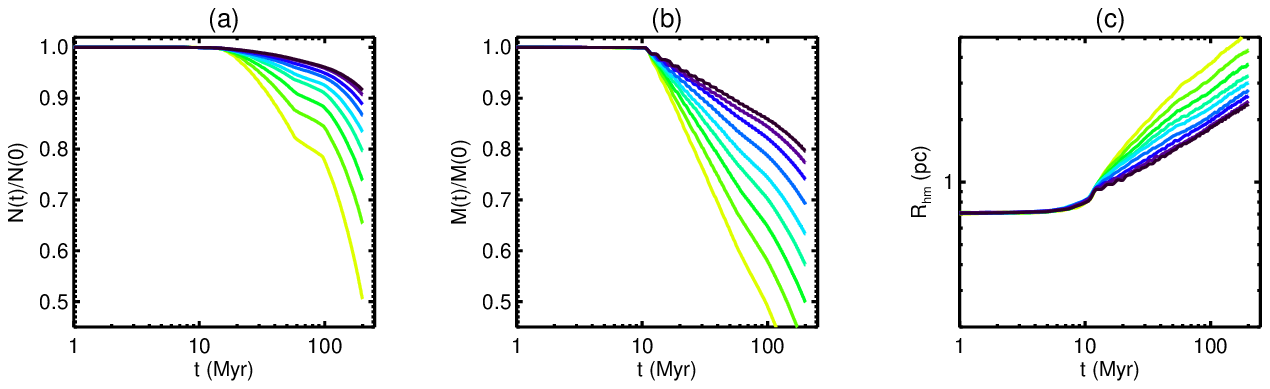}\\
  \includegraphics[width=1\textwidth,height=!]{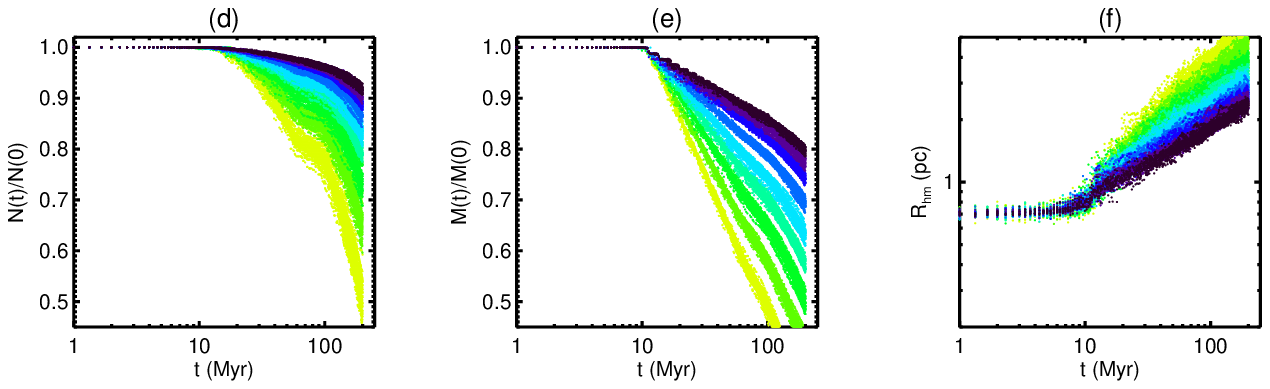}
  \caption{As Fig.~\ref{fig:mn}, but for the star clusters in model~MR with stellar evolution ({\em top panels}). The bottom panels show the spread among the individual models in our ensemble of realisations.
  \label{fig:me} }
\end{figure*}

\begin{figure}
  \centering
  \includegraphics[width=0.4\textwidth,height=!]{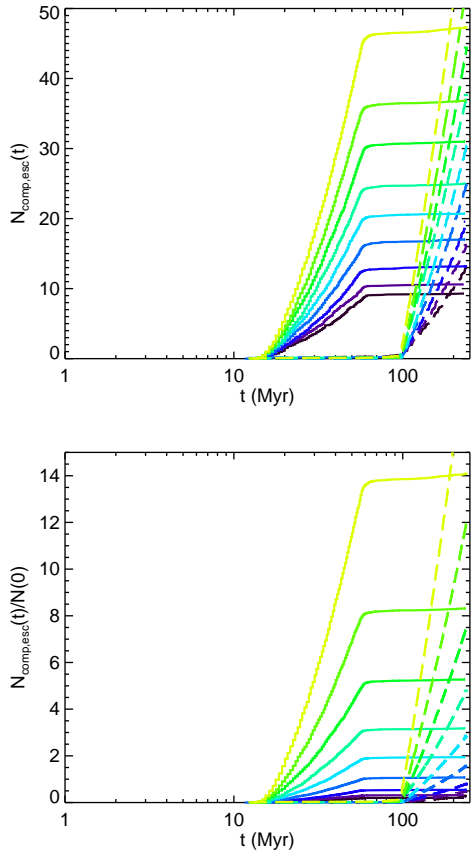}
  \caption{Escaping compact objects for model~MR with stellar evolution (cf. Fig.~\ref{fig:mn}). The top panel shows the {\em cumulative} number of compact objects $N_{\rm comp,esc}$ that have escaped over time. The bottom panel shows $N_{\rm comp,esc}/N(0)$, where $N(0)$ is the initial number of stars (of all masses) in the star cluster (see Table~\ref{table:models}). Solid and dashed curves indicate the results for neutron stars and white dwarfs, respectively. All curves are averages for the ensemble of simulations. In each panel
    the darkest curves are for IMF low-mass cut-offs of $0.1\msun$,
    becoming lighter as the mass of the low-mass cut-off increases to $2\msun$.
  \label{fig:neutron} }
\end{figure}
\begin{figure}
  \centering
  \includegraphics[width=0.4\textwidth,height=!]{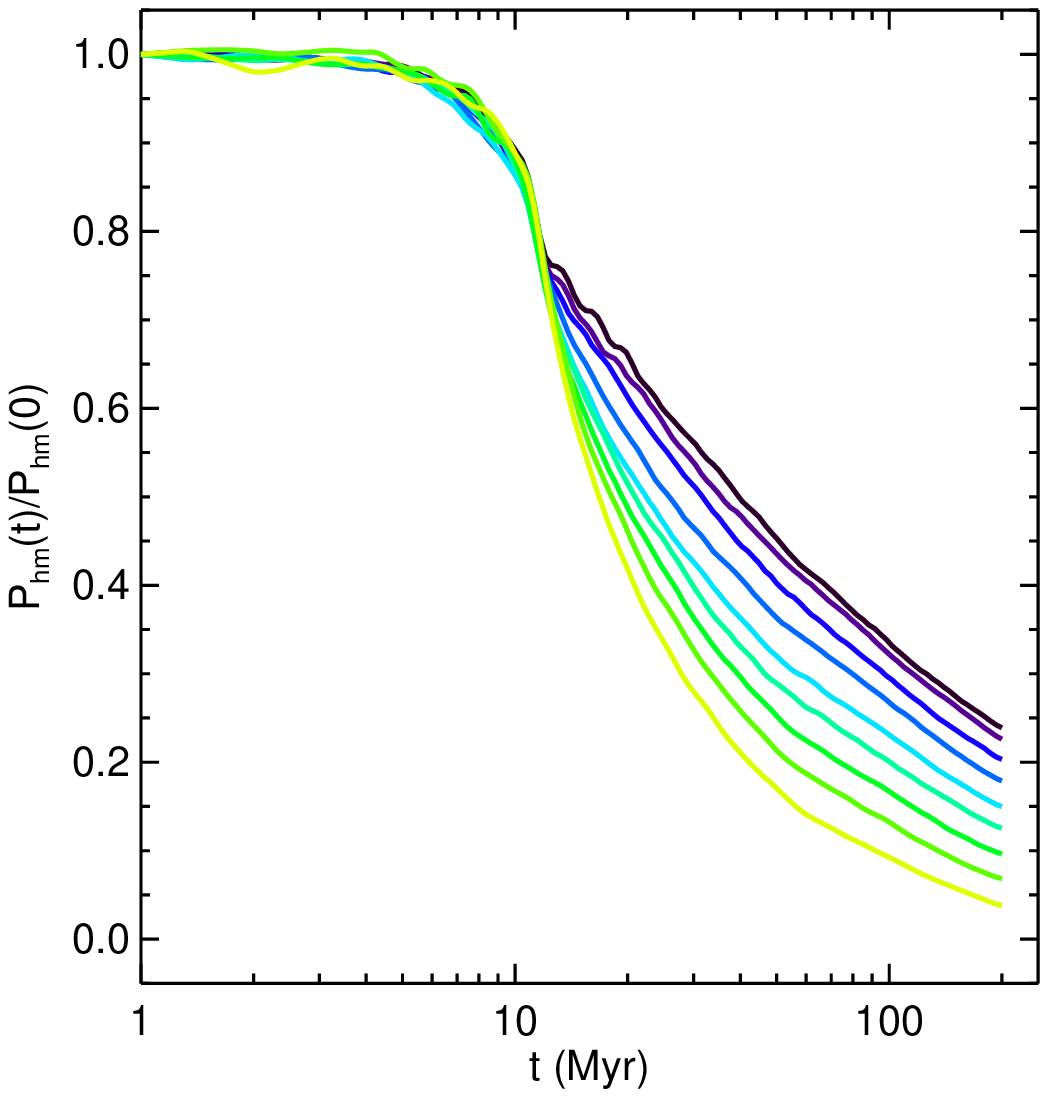}
  \caption{The evolution of the normalised gravitational potential $P_{\rm hm}(t)/P_{\rm hm}(0)$ at the half-mass $\halfmass$ radius for model~MR with stellar evolution. All curves are averages for the ensemble of simulations. In each panel
    the darkest curves are for IMF low-mass cut-offs of $0.1\msun$,
    becoming lighter as the mass of the low-mass cut-off increases to $2\msun$.
  \label{fig:potential} }
\end{figure}


First we consider model~MR.  In this model we keep $\mcl$ and $\halfmass$ constant.
These could be considered the models in which the most basic cluster
parameters are kept the same and might be argued to be those in which
the clusters are truly `the same'.  Note that the initial tidal
radius is the same for each cluster we consider here.

In the MR models the initial cluster masses are always
$\mcl=1500\msun$, and the initial half-mass radii are
$\halfmass=0.59$~pc.  Since $\mcut$ changes from $0.1\msun$ to $2\msun$,
$N$ decreases from $N=4607$ ($\avmass = 0.33\msun$) to $N=336$
($\avmass = 4.5\msun$).  As $\mcl$ and $\halfmass$ are initially identical for each cluster, so are
the crossing times.  But as $N$ decreases, the
relaxation time falls from 20~Myr ($\mcut = 0.1\msun$) to 2~Myr ($\mcut = 2\msun$).

For model~MR we first consider simulations with no stellar evolution,
shown in Fig.~\ref{fig:mn}.  This paper contains several very similar
figures, so it is worth describing them in some detail.  Each panel
contains several curves with different colours.  Darker shades
show lower values of $\mcut$ from $0.1\msun$ (darkest colour) to $2\msun$ (lightest colour).  In each figure, panel~(a)
shows the evolution of the relative numbers of stars in each cluster
with time.  Panel~(b) shows the evolution of the
relative mass with time and panel~(c) shows the
evolution of the half-mass radii with time. All results represent the average of an ensemble of simulations.

When we ignore stellar evolution as we do in Fig.~\ref{fig:mn} the
evolution of clusters will be entirely a result of
dynamics.  One might expect two-body relaxation to dominate. This
would mean that $\mcut = 2\msun$ clusters should `evolve' around 10 times faster
than  $\mcut = 0.1\msun$ clusters (see Eq.~\ref{eq:relaxationtime}).  By `evolve' we mean that the rates at which energy equipartition and ejections occur should be 10 times
faster, but this will also be moderated by contraction of the core and expansion of the half-mass radius in response to ejections and evaporation (note that ejections can occur with positive energy owing to the tidal truncation).

However, examination of Fig.~\ref{fig:mn} shows that the evolution of
$\mcut = 0.1\msun$ clusters (darkest colour) and $\mcut = 2\msun$
clusters (lightest colour) are really quite similar and a significant
expansion occurs in all cases.  Higher-$\mcut$ clusters
evolve {\em slightly} faster (the lightest colour curves are offset), but
the difference is not as significant as one might have expected, and
certainly not a factor of 10 in the `speed' of the evolution.

The reason for this is that in all of these clusters the dynamics is
actually dominated by a massive central binary. The dynamical formation of a massive binary system in the centre of a star cluster is very common, and this binary is
made of two of the most massive stars in the cluster \citep[see, e.g.,][]{aarseth2003, heggiehut}. It acts
to heat the cluster, causing the expansion seen in the half-mass radius
(Fig.~\ref{fig:mn}(c)). This heating occurs on the time-scale of the crossing time,
which is the same in each of these clusters.
This shows that dynamics is not as simple as two-body relaxation and
can be driven on different time-scales if binaries are present.  

The situation we have simulated in Fig.~\ref{fig:mn} is not
particularly physical. Apart from the fact that stellar evolution occurs in reality,  stellar evolution will be especially important
for two other reasons.  First, it will cause mass loss and drive
expansion, and second, it will also evolve the most massive stars, which are
the stars that tend to be in the dynamically important binary which
drives the evolution. 

In Fig.~\ref{fig:me} we show the results of constant $\mcl$ and
$\halfmass$ clusters {\em with} stellar evolution.  Here we have the
basic result that one would expect: clusters with a higher-$\mcut$ 
lose more mass more rapidly, expand more rapidly and are
destroyed more rapidly. In addition, we show in the bottom panels of Fig.~\ref{fig:me} the results for the individual models in the ensemble of simulations, which give an indication of the spread resulting from stochasticity in the initial conditions and chaos afterwards.

What is particularly interesting is not that clusters with a higher $\mcut$
lose mass more rapidly (they cannot {\em not} lose more mass), but that their mass
loss is not as dramatic as one might have expected.
In the very extreme case of a cluster with $\mcut = 2\msun$ there is still a surviving cluster after 200~Myr.  And the
evolutionary sequences of clusters with $\mcut = 0.1$ to $0.5\msun$ are very similar.

Fig.~\ref{fig:me}(b) and Table~\ref{table:results} show that this mass loss is not caused only by stellar evolution.  The star clusters with $\mcut = 0.1\msun$ and 
$\mcut = 2\msun$ lose 15 and 55 per
cent of their mass, respectively, in 200~Myr, only owing to stellar evolution.
However, simulations including dynamics show that the mass loss is 20
per cent in the $\mcut = 0.1\msun$ clusters, and 72 per cent in the
$\mcut = 2\msun$ clusters.  

In all cases the mass loss is dominated
by the mass loss as a result of stellar evolution, and ejections/tidal
overflow only account for about a quarter to a third of the
mass loss.  

It seems unexpected that the contribution of dynamical
mass loss is very similar in all cases.  Not only do high-$\mcut$
clusters lose more mass through stellar evolution, but their initial
relaxation times are much shorter, which would be expected to drive
faster dynamical evolution as well.

However, the rapid and significant mass loss due to stellar evolution
causes high-$\mcut$ clusters to expand significantly.  This expansion
increases their crossing times, and so significantly reduces their
relaxation times, thus `slowing' their dynamical evolution.

In Fig.~\ref{fig:me}(c) we see that the half-mass radii of
high-$\mcut$ clusters (lighter colour curves) expand by factors of
several.  Indeed, all clusters expand from their initial half-mass
radii of 0.6~pc to between 2.4~pc ($\mcut = 0.1\msun$) to 5.3~pc
($\mcut = 2\msun$).  Therefore, all clusters have significantly longer
relaxation times after 200~Myr than their initial values.

Crossing times scale as $\halfmass^{3/2} \mcl^{-1/2}$, so for the
$\mcut = 0.1\msun$ clusters the crossing time has increased by a factor of
approximately 10 after 200~Myr, but for the $\mcut = 2\msun$ clusters it has
increased by a factor of 50.  This acts to help equalise the initial
difference in which the initial relaxation times of the $\mcut = 2\msun$ clusters was 10 times shorter (changing $N$ also plays a role
here to decrease the relaxation times of the $\mcut = 2\msun$ clusters
more, but it is less significant).

Binary heating can also play a minor role.  Fig.~\ref{fig:me}(c) shows
that low-$\mcut$ clusters keep expanding significantly, even at late
times when stellar evolutionary mass loss becomes small
(especially after 100~Myr).  This expansion is driven by binary
heating (although not to the extent that clusters without
stellar evolution because of the lower mass of the binaries -- compare
Figs~\ref{fig:mn}(c) and~\ref{fig:me}(c)).  

An interesting feature is present in the evolution of the number of
stars in the clusters in Fig.~\ref{fig:me}(a).  The number of stars in
the high-$\mcut$ clusters falls sharply during the first $10-60$~Myr, and
then slows significantly before declining rapidly again after about
100~Myr.  This feature is most prominent in the highest-$\mcut$
clusters, reducing in importance as smaller values for $\mcut$ are chosen.

The same feature is present in the fractional mass loss shown in
Fig.~\ref{fig:me}(b), but to a much lesser extent (a slight change in
the slope of the mass-loss line for high $\mcut$).
This feature is the result of the supernovae of relatively large
numbers of stars.  The immediate effect of stellar evolutionary mass
loss is to reduce the mass of
the cluster, but not the number of stars in the cluster: massive
stars change from being $10-20\msun$ stars into being $\approx 1.4\msun$
neutron stars.  This has two effects.  

First, neutron stars are given velocity kicks
\citep[for details see][and \texttt{NBODY6}]{aarseth2003}
 which most often leads to them being ejected
from the cluster -- this causes the number of stars to fall fairly
rapidly. The production and rapid escape of neutron stars halts around 60~Myr, which explains the kink in Fig.~\ref{fig:me}(a) at this time. Lower-mass stars that do not go supernovae evolve into white dwarfs. These white dwarfs do not get a high-velocity kick, and only escape at around 100~Myr, which explains the second kink in Fig.~\ref{fig:me}(a).

Second, the very significant mass loss caused by stellar evolution
unbinds a high-velocity tail of stars in the initial velocity
distribution.  These newly unbound stars take some time to escape the
cluster and so are associated with the cluster for some time.  This is
an effect very similar to that seen in simulations of gas expulsion
from star clusters \citep[see especially][who detail
the `luminosity bump' caused by slow escapers]{bastian2006}.

Therefore, we have three effects that cause the {\em number} of stars to
decrease.  First, velocity kicks on neutron stars which are responsible for
low-mass objects to be lost. This is clearly seen in Fig.~\ref{fig:neutron}, which shows the cumulative number of neutron stars that have escaped the star clusters over time. In fact, only $1-5$ per cent of the stars with masses larger than $10\msun$ escape before they evolve, while all others experience mass loss while still being a member of the star cluster. Second, the number of stars decreases following 
the unbinding of high-velocity
stars due to the change in the gravitational potential from stellar evolutionary
mass loss and the resulting expansion of the star cluster. The potential $P_{\rm hm}(t)$ at the half-mass radius, shown Fig.~\ref{fig:potential}, exhibits a strong decrease around 10~Myr, which causes part of the stellar population (which at $t=0$~Myr has the same velocity dispersion in all models~MR) to escape, and this is most pronounced for the clusters with a high $\mcut$.  Third, the `normal' process of two-body encounters and
ejections.  All of these processes lead to the loss of relatively
low-mass stars in clusters with high $\mcut$, leading to a different
rate of change in mass loss and number loss.

Therefore, in Fig.~\ref{fig:me}(a) we see significant loss by number during the
first $10-60$~Myr as a result of the violent early evolution. Subsequently, there is a slowing of loss by number at $60-100$~Myr once all the fast stars have passed
over the tidal boundary and are `lost' by the cluster.  Then a
speeding up of the loss of stars after around 100~Myr as the cluster starts
to fill (its now smaller) tidal radius.

An interesting aside with observational consequences is that 
the average mass of a star in any cluster remains roughly
constant (to within a factor of two) after around 20~Myr.  The most
massive stars evolve, but lower-mass stars are ejected, causing only a
very gradual decline in the average stellar mass in a cluster with
time.  

Changing $\mcut$ means that even though clusters in the MR models
start at the same mass, their luminosities will be
very different.  Initially, clusters with a high $\mcut$ will be {\em
  much} more luminous than those with a low $\mcut$.  It might be
thought that, as the high-$\mcut$ clusters lose so much more of their
mass,  they will become less luminous.

Let us take our two extreme values of $\mcut$ after 200~Myr.  Starting
from initially $1500\msun$ clusters, the $\mcut = 0.1\msun$ clusters
have become $1200\msun$ clusters with a mean stellar mass of
approximately $0.7\msun$.  The $\mcut = 2\msun$ clusters have declined in
mass to only $\sim 400\msun$, but the mean mass of a star is $2\msun$.  So whilst the $\mcut = 0.1\msun$ clusters have more than 10
times more stars remaining, the stars that remain in the $\mcut = 2\msun$ clusters are around 40 times more luminous.  Therefore, even
though the high-$\mcut$ clusters have lost much more of their mass,
they are still more luminous than the low-$\mcut$ clusters.

In summary, for clusters with
the same initial mass and initial (half-mass) radius, those with a
higher-$\mcut$ do lose more mass.  But this is only
really significant in clusters with extremely high values of $\mcut$
($> 1\msun$).  For low values of $\mcut$, the evolution of different
clusters is actually very similar, this is due to expansion slowing
dynamical evolution and the equal importance of binary heating in
different clusters.  This results in a roughly equal importance and
rate of dynamical mass loss in all clusters, regardless of $\mcut$.

Comparisons of equal-mass and equal-size clusters
appear the most sensible, but these would observationally be quite different.  If the
low-mass component is invisible (e.g. because of distance), then the
clusters with a higher-$\mcut$ would appear much more luminous
(since more of their mass would be in higher-mass stars).  Therefore,
 saying what constitutes `the same' is difficult.



\subsection{Identical $\trelax$ and $\halfmass$ (model~TR)} \label{section:modelt}

\begin{figure*}
  \centering
  	    \includegraphics[width=1\textwidth,height=!]{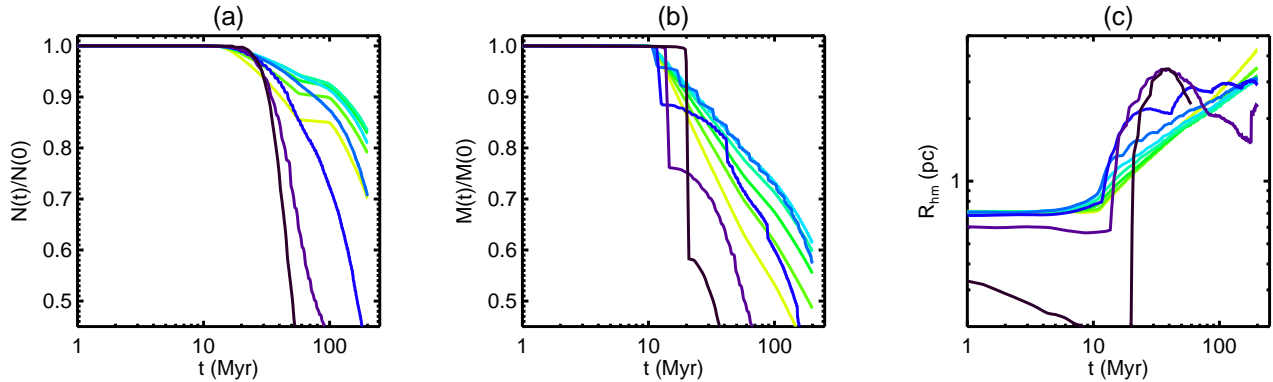}
	\caption{As Fig.~\ref{fig:mn}, but for the star clusters in model~TR with stellar evolution. 
    \label{fig:te} }
\end{figure*}


Another way of making clusters `the same' is to compare clusters which 
initially have the same (dynamical) evolutionary time-scale, i.e., the same initial
relaxation time $\trelax$.  If the initial relaxation times of all
clusters are the same, then we are considering clusters with the same
{\em initial} dynamical time-scales -- therefore differences should be solely due to
different stellar evolutionary mass loss.

Identical $\trelax$ can be obtained
using Eq.~(\ref{eq:relaxationtime}) with a fixed $\halfmass$ but varying
the number of stars, $N$, and the total mass of the cluster, $\mcl$.
Because $\mcl$ depends on both $N$ and $\mcut$ this is slightly
non-trivial. In the second block (model~TR) of Table~\ref{table:models} we can see how
both $\mcl$ and $N$ must vary with $\mcut$ in order to keep a fixed
initial $\trelax = 5$~Myr.  

The potential problem with these models is clear when we compare the
different $N$ needed in different clusters to keep $\trelax$ constant
for a constant $\halfmass$.  When $\mcut = 2\msun$,  $N=3961$, a
reasonably large number.  However, when $\mcut = 0.1\msun$ we require
$N=82$ which is so low we would expect the evolution to be driven by
stochastic encounters rather than any statistically `smooth' evolution.
Indeed, $\mcut = 0.1\msun$ simulations are so stochastic that we do
not illustrate them.

Also note that, by changing the initial cluster mass, we
also increase the tidal radius by a factor of 8 between $\mcut = 0.1$
and $2\msun$. This allows the clusters with higher $\mcut$ to expand more, which decreases the time-scale at which they evolve dynamically.

The final results of simulations after 200~Myr with 
stellar evolution are again listed in Table~\ref{table:results}, and their evolution
shown in Fig.~\ref{fig:te}.  Interestingly, the evolution with $\mcut$
is not simple and falls into two `regions'.

In all panels of Fig.~\ref{fig:te} the low-$\mcut$ clusters (darker colour curves) show rapid and stochastic evolution due to their small $N$
and low mass (and therefore small tidal radii).  All
evolve very rapidly and can form binaries that dominate their
evolution (in the models with the smallest $\mcut$ the core radius falls dramatically in a
core-collapse-type event before the star cluster is blown apart).  It is difficult
to draw any conclusions about evolutionary trends in the low-$\mcut$
cases because $N$ is so low.  It is also unclear if such low-$N$ objects
constitute a `cluster' under any sensible definition \citep[see, e.g.,][for a discussion on this topic]{gieles2011}.

In cases of high-$\mcut$ (with reasonable $N$) we would expect much
less stochastic evolution and this is what we see.  Given that each of the clusters
has the same initial relaxation time, we might expect stellar evolution
to dominate over all else, and this appears to be the case.

Fig.~\ref{fig:te}(c) shows that each of the high-$\mcut$ clusters
expands by approximately the same fraction and at roughly the same
rate.  This means that higher-$\mcut$ clusters do not significantly
increase their relaxation times relative to low-$\mcut$ clusters
(although the total mass plays a role).  

The expansion is significant enough that in all cases clusters are
starting to fill their tidal radii by $100-200$~Myr and it is tidal
overflow that dominates their mass loss at late times.  This effect is
of roughly equal importance in all the high-$N$ clusters since even
though the high-$\mcut$ clusters have lost relatively more mass (thus reducing
their tidal radii), they were initially more massive and so started
with larger tidal radii.

To summarise, in the case where we keep the relaxation time and the
half-mass radius constant, we get the rather unexpected result that the
clusters that survive the longest have intermediate $\mcut$, and
those which are destroyed most rapidly have the lowest $\mcut$.  

However, at the low-$\mcut$ end this is due to
low-$N$ stochastic effects in the dynamics.  At the high-$\mcut$ end
this is due to the significant differences in cluster mass that we
need to keep the relaxation times constant, leading to very different
tidal boundaries.  

We would also argue that nobody would sensibly 
describe two clusters as being `the same' if their masses differ by
several orders of magnitude.



\subsection{Identical $\mcl$ and $\trelax$ (model~MT)} \label{section:modelr}

\begin{figure*}
  \centering
    \includegraphics[width=1\textwidth,height=!]{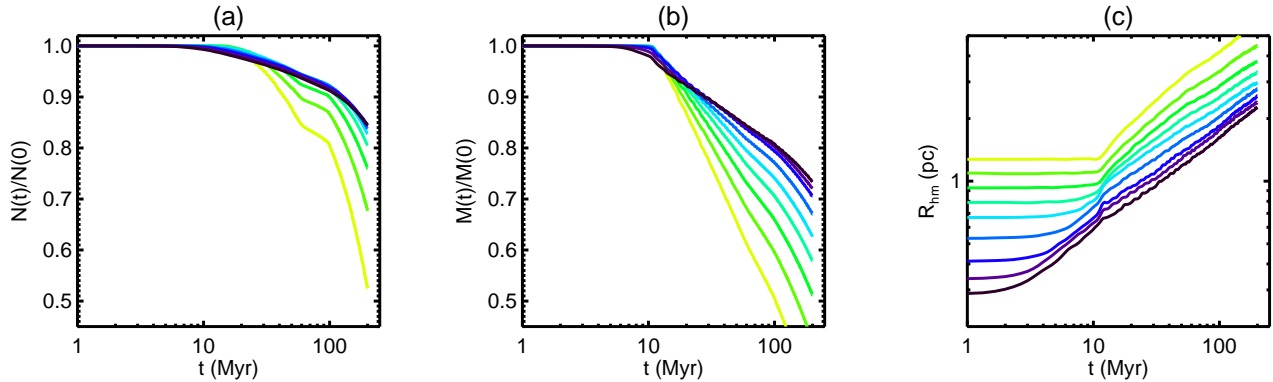}
    \caption{As Fig.~\ref{fig:mn}, but for the star clusters in model~MT with stellar evolution. 
    \label{fig:re} }
\end{figure*}

In order to avoid the problems introduced by low-$N$ stochasticity and
large differences in tidal radii in the
constant $\trelax$ and $\halfmass$ models above we can instead keep
$\mcl$ and $\trelax$ constant and vary $N$ and $\halfmass$.
In the third block of Table~\ref{table:models}
(model~MT) we can see that to keep $\mcl$ and $\trelax$ constant we
need to vary between $N=4627$ and $\halfmass = 0.24$~pc for $\mcut =
0.1\msun$, and $N=338$ and $\halfmass = 1.06$~pc for $\mcut = 2\msun$.  
Note that this is the opposite trend in $N$ with $\mcut$
as previously, now as $\mcut$ increases, $N$ decreases.

Here clusters are not too `different' -- their masses are the same,
and their radii only differ by a factor of approximately four, although 
the number of stars in each cluster can vary by a factor
of over 10.  Again we summarise the final states in Table~\ref{table:results}, and
show the evolution of the key cluster parameters in Fig.~\ref{fig:re}.

In Fig.~\ref{fig:re} we again see the expected trend that clusters with
higher $\mcut$ and in which stellar evolution is more important
(lighter colour curves) lose more mass than low-$\mcut$ clusters.
But yet again, the differences in the evolution of clusters with different $\mcut$ is
 not as extreme as one might expect.  By 200~Myr clusters
with $\mcut = 0.1$ and $1\msun$ have only lost between 27 and 49 per
cent of their initial mass, respectively -- not a great difference for
two such different low-mass cut-offs, both about twice that 
expected from stellar evolutionary mass loss alone (15 and 28 per
cent, respectively).

The reason that the differences are not as great as one might expect
is that the evolution of the low-$\mcut$ clusters is driven by binary
heating due to a dense initial state.  Low-$\mcut$ clusters have more
stars per unit mass, so in order to keep $\trelax$ constant,
$\halfmass$ must be much smaller.  For $\mcut = 0.1\msun$, $\halfmass
= 0.24$~pc initially, compared to $0.77$~pc when $\mcut = 1\msun$.
This leads to central densities of $6 \times 10^4$ stars pc$^{-3}$ in the
$\mcut = 0.1\msun$ clusters (compared to approximately 50 stars pc$^{-3}$ in the
$\mcut = 2$ clusters).
As can be seen in Fig.~\ref{fig:re}(c), the $\mcut = 0.1\msun$
clusters start expanding immediately (and before stellar evolution has
any effect).  This is caused by rapid binary formation and heating.

Therefore, even though all clusters
have the same $\trelax$ initially, within just 10 Myr the $\mcut = 0.1\msun$ clusters have expanded by a factor of approximately four, decreasing
their relaxation times and `slowing' their dynamical evolution. The star clusters with $\mcut = 2\msun$, on the other hand, do not experience any significant expansion during the first 10~Myr, and lose relatively fewer stars and less mass during this time than the clusters with smaller $\mcut$. Beyond 10~Myr stellar evolution sets in, driving expansion, and loss of stars and mass, which is particularly important for the clusters with high-$\mcut$, such that they overtake those with lower $\mcut$ in terms of mass loss.
By 200~Myr, the $\mcut = 0.1\msun$ clusters have expanded by a factor
of 10, compared to a factor of 5 for the $\mcut = 1\msun$ clusters
(in their case driven mainly by stellar evolutionary mass loss).

Thus, even though all these clusters have the same initial relaxation
time-scales, the low-$\mcut$ clusters change so rapidly that this initial similarity
disappears almost immediately.  One could stop the low-$\mcut$
clusters evolving so rapidly by increasing their half-mass radii by a
factor of, say, 10.  However, the same scaling would have to be
applied to the high-$\mcut$ clusters, giving `clusters' of a few
hundred stars with half-mass radii of 10~pc.  Even if such a `cluster'
were formally bound at formation, tidal forces would soon destroy it.

In summary, it is possible to construct initial conditions that have the
same cluster mass and relaxation time, but the required differences in radii would
either drive rapid dynamical evolution in extremely dense clusters, or
tidal forces would destroy extremely low-density clusters.  Therefore,
clusters that start `the same' cannot remain so for long because of
processes that have nothing to do with stellar evolutionary mass loss.



\subsection{An identical upper IMF (model~UR)} \label{section:modelk}

\begin{figure*}
  \centering
    \includegraphics[width=1\textwidth,height=!]{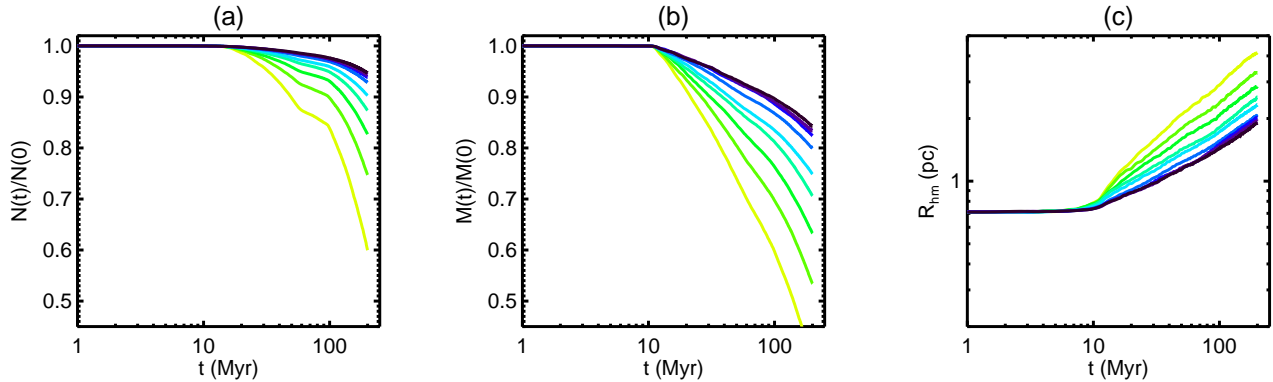}
    \caption{As Fig.~\ref{fig:mn}, but for the star clusters in model~UR with stellar evolution.
    \label{fig:he} }
\end{figure*}

Finally, we describe a set of models in which the high-mass stellar
content and the half-mass radii of each cluster are the same.  By this
we mean that there are the same number of stars with masses greater
than $3\msun$ in every cluster: such clusters
would appear very similar to observers if seen at a significantly large
distance that the low-mass population is `invisible'.

If the high-mass stellar content is the same, then for high $\mcut$
there will be few other stars in the cluster, but for low $\mcut$
there will be a significant low-mass population.  We use the \cite{kroupa2001}
IMF in the mass range $\mcut < M < 20\msun$ and change $\mcut$.  In
all cases we ensure that there are exactly 200 stars with masses $3 <
M < 20\msun$.

This means that for $\mcut = 2\msun$ the total number of stars is
only 352, and the total mass apporximately $1600\msun$.  But for $\mcut = 
0.1\msun$ the total number of stars is
8263, and the total mass approximately $4700\msun$.  This means that the
relaxation times vary from 2~Myr for $\mcut = 2\msun$ to 19~Myr for $\mcut = 
0.1\msun$.  The properties of these models are listed in
Table~\ref{table:models}.  The results of the simulations are shown in
Fig.~\ref{fig:he}.

The high-$\mcut$ clusters must lose much more mass via stellar
evolution 
than the low-$\mcut$ clusters, and they have much shorter relaxation times, and are
less massive and so have smaller tidal radii.  Given all of this, one
would expect to see much more rapid and significant mass loss from
high-$\mcut$ clusters.

There is a trend to greater mass loss from higher-$\mcut$ clusters
(especially in the extreme $2\msun$ cut-off), but it is probably not
as great as one would expect.  From previous arguments one can see why
this is the case. 

First, low-$\mcut$ clusters evolve `faster' than one would
think. Fig.~\ref{fig:he}(c) shows the rapid expansion of low-$\mcut$
clusters (as we saw earlier in Fig.~\ref{fig:re}(c)).  This is driven
by the high number densities in these clusters causing binary
formation and this driving expansion and ejections. Although difficult to see in
Fig.~\ref{fig:he}(a) and~(b), a handfull of stars are ejected before
10~Myr for models with small $\mcut$ but not for those with large $\mcut$, and this difference cannot possibly be due to stellar evolution.
The huge expansion of the low-$\mcut$ clusters decreases their
relaxation times, but the binaries formed in the dense phase are
efficient at ejecting stars and keep heating the clusters.

Also, the high-$\mcut$ clusters evolve more `slowly' than their
initial relaxation times would lead one to believe because the
significant early stellar evolutionary mass loss drives expansion and
significantly decreases the relaxation times.



\section{Discussion} \label{section:discussion}

A key evolutionary property of any star cluster is the rate at which
it loses mass, and
on what time-scale it is destroyed.  Only by knowing these things can
we start to `reverse engineer' observed clusters and populations of
clusters to their initial states.

The evolution of a star cluster is determined by two fundamental
physical processes.  First, by dynamics: stars will be ejected over the
tidal boundary due to internal dynamics or external perturbations.
Second, by stellar evolution: stars evolve and lose mass, so
reducing the mass of the cluster.  But as we have seen, stellar
evolution causes changes in the radius and energy of the cluster, which
change dynamical time-scales, so these processes are not independent.

In this paper we have investigated the effect of variations in the
low-mass cut-off of the IMF on the evolution of clusters.  We have
simulated clusters in which the low-mass cut-off of the IMF varies
between $0.1\msun$ and $2\msun$.  For most cases we have used a pure
Salpeter power law over all masses which is not a particularly
realistic IMF, but serves our purposes allowing for simple tests of
the general effects of altering the IMF.

The expectation of what will happen when the IMF is varied (and these
were certainly our expectations on starting this project) is that
a higher $\mcut$ in the IMF will cause clusters to lose more mass
more rapidly and be destroyed much more rapidly.  One would expect
this as a cluster with a cut-off at $0.1\msun$ will lose around 15 per
cent of its mass in 200~Myr as a result of stellar evolution alone, while a
cluster with a cut-off at $2\msun$ will lose around 55 per cent.

A completely unsurprising result is that clusters with higher-mass 
cut-offs in their IMFs lose more mass.  This is unavoidable.
However, the results we have described above show that changing the
IMF has many important, but rather subtle, effects beyond simply
altering the total amount of mass lost by stellar evolution and thus
speeding destruction.

{\em What is `the same'?}  An important, but subtle, effect of changing the
IMF is that it becomes very unclear what should be compared to what to
determine the relative `speed' of evolution.
Changing the IMF means changing the number of stars per unit mass.  An
IMF with a high-mass cut-off has few stars per unit mass, while a 
low-mass cut-off has many more stars per unit mass.  Therefore
clusters with the same total mass have very different numbers of
stars.  
Thus in order to have any two of mass, radius, and relaxation time kept
constant, the third property must be changed significantly between
different cut-offs (sometimes by orders of magnitude).
We would argue that it is impossible and unphysical to ever have two
clusters with different cut-offs in the IMF that can otherwise be
described as `the same'.
This is true in terms of the physical properties of a
cluster (mass, radius, relaxation time).  But it also impacts on the
observable properties of a cluster such as the colour/luminosity
evolution because the mass-to-light ratio evolves in very different ways.
We do not consider these observation problems any further in this
paper and concentrate on the underlying physical properties.

{\em How does stellar evolution impact dynamical evolution?}  Stellar
evolution and dynamical evolution cannot be separated from one
another.  Stellar evolution causes mass loss from a cluster which
alters the mass and energy of the cluster.  In responding to this
clusters will expand, which will increase their crossing times, and
hence their relaxation times.

Clusters with a high-mass cut-off in their IMF will lose more mass
through stellar evolution than those with a low-mass cut-off.  
Therefore they will expand more, and reduce their relaxation times
more, and hence dynamically evolve more `slowly'.  The effect of this
is to allow clusters with high-mass cut-offs to survive longer than
one might think looking at their initial conditions.
The interplay of these effects is rather complex since it depends on what
two properties of the cluster were initially `the same'.  

If a cluster initially has a large radius then greater
expansion can push a cluster towards overflowing its tidal limit thus
greatly speeding up its destruction.  Alternatively, clusters with
very small initial radii can evolve significantly before stellar
evolution becomes important because of binary formation and heating.


\subsection{Future work}

This rather idealised work has shown that the effects of changing the
IMF on the evolution of star clusters are rather more subtle than one
might have thought.  Several interesting lines of inquiry could be
followed from this work.  

Our treatment of the IMF is rather simple, and a more
comprehensive study should consider changes to both low- and high-mass
cut-offs as well as to the shape of the IMF.

A dynamical effect that can be very important is heating by
binaries.  In a number of simulations with very high initial central
densities ($10^4$ to $10^5$ stars pc$^{-3}$) binaries can form which
heat the cluster causing expansion and increased ejections (because of the
larger cross section of a binary).  Our simulations contain no 
primordial binaries, despite observational evidence that large fraction of stars in the Galactic field \citep[e.g.,][]{raghavan2010}, in young associations \citep[e.g.,][]{kouwenhoven2005, kouwenhoven2007}, and in star-forming regions \citep[e.g.,][]{connelley2008, duchene2013, chen2013, reipurth2014} are part of a binary or multiple system. It is therefore interesting and necessary to include them to investigate their effect(s).

We have touched upon the implications to the observed properties of
clusters with different cut-offs in their IMFs but this is clearly
something of great importance if one believes that the IMF does vary in
some environments.



\section{Conclusion} \label{section:conclusion}

We have performed simulations of ensembles of clusters characterised by 
different mass cut-offs at the bottom of the IMF.

Our main result is that it is impossible to compare the effects of
altering the IMF on two clusters that are otherwise `the same.'  If
the IMF is changed then this changes the number of stars per unit
mass, therefore only any two (or some complex combination of) mass,
radius, and relaxation time can ever be `the same' initially.

As well as not being `the same' initially, different cut-offs in the
IMF cause clusters to evolve differently.  Clusters with more stellar
evolutionary mass loss will expand more and hence increase their
relaxation times and `slow' their dynamical evolution.

In conclusion, the effect of varying the IMF on the evolution of star
clusters is rather subtle and complex.  Star clusters that contain many
more high-mass stars do lose more mass due to stellar evolution, but
the impact this has on their destruction might not be as great as one
might naively expect.


\section*{Acknowledgments}

We thank the anonymous referee for her/his insightful comments that helped to improve this paper.
MBNK was supported by the Peter and Patricia Gruber Foundation through the PPGF fellowship, by the Peking University One Hundred Talent Fund (985) programme, by the National Natural Science Foundation of China (grants 11010237, 11050110414, 11173004), and by STFC under grant number PP/D002036/1. This publication was made possible through the support of a grant from the John Templeton Foundation and the National Astronomical Observatories of the Chinese Academy of Sciences. The opinions expressed in this publication are those of the author(s) do not necessarily reflect the views of the John Templeton Foundation or the National Astronomical Observatories of the Chinese Academy of Sciences. The funds from the John Templeton Foundation were awarded in a grant to The University of Chicago which also managed the programme in conjunction with the National Astronomical Observatories, Chinese Academy of Sciences.  
RdG acknowledges partial research support through grants 11043006, 11073001 and 11373010 from the National Natural Science Foundation of China.
We thank the  British Council for networking funding through a `Research Co-operation grant' between the University of Sheffield and Kyung Hee University under the Prime Minister's Initiative-2 (PMI2) programme.
MR acknowledges funding from the Nuffield Foundation for a 2008 Undergraduate Summer Research Bursary, URB/35327, and STFC (grant number ST/G001758/1). 
SSKÕs work was supported by the Mid-career Research Program
(No. 2011-0016898) through the National Research Foundation (NRF)
grant funded by the Ministry of Education, Science, and Technology (MEST)
of Korea.



\bsp
\label{lastpage}

\begin{thebibliography}{}

\bibitem[Aarseth(1973)]{aarseth1973} Aarseth, S.~J.\ 1973, Vistas in Astronomy, 15, 13 

\bibitem[Aarseth et 
al.(1974)]{aarseth1974} Aarseth, S.~J., Henon, M., \& Wielen, R.\ 1974, \aap, 37, 183 

\bibitem[Aarseth(2003)]{aarseth2003} Aarseth, S.~J.\ 2003, 
Gravitational N-Body Simulations, pp.~430.~ISBN 
0521432723.~Cambridge, UK: Cambridge University Press

\bibitem[Allison et al.(2010)]{allison2010} Allison, R.~J., 
Goodwin, S.~P., Parker, R.~J., Portegies Zwart, S.~F., 
\& de Grijs, R.\ 2010, \mnras, 407, 1098 

\bibitem[Barnab{\`e} et al.(2013)]{barnabe2013} Barnab{\`e}, M., 
Spiniello, C., Koopmans, L.~V.~E., et al.\ 2013, \mnras, 436, 253 

\bibitem[Bastian 
\& Goodwin(2006)]{bastian2006} Bastian, N., \& Goodwin, S.~P.\ 2006, \mnras, 369, L9 

\bibitem[Bastian et al.(2007)]{bastian2007} Bastian, N., 
Konstantopoulos, I., Smith, L.~J., Trancho, G., Westmoquette, M.~S., 
\& Gallagher, J.~S.\ 2007, \mnras, 379, 1333 

\bibitem[Bastian et 
al.(2010)]{bastian2010} Bastian, N., Covey, K.~R., \& Meyer, M.~R.\ 2010, \araa, 48, 339 

\bibitem[Bastian et al.(2012)]{bastian2012} Bastian, N., Adamo, A., 
Gieles, M., et al.\ 2012, \mnras, 419, 2606 

\bibitem[Baumgardt \& Makino(2003)]{baumgardt2003} Baumgardt, H., \&
Makino, J.\ 2003, \mnras, 340, 227

\bibitem[Baumgardt et al.(2013)]{baumgardt2013} Baumgardt, H., 
Parmentier, G., Anders, P., \& Grebel, E.~K.\ 2013, \mnras, 430, 676 

\bibitem[Bekki(2013)]{bekki2013} Bekki, K.\ 2013, \apj, 779, 9 

\bibitem[Binney \& Tremaine(1987)]{binneytremaine} Binney, J., \&
Tremaine, S.\ 1987, Princeton, NJ, Princeton University Press

\bibitem[Brewer et al.(2012)]{brewer2012} Brewer, B.~J., Dutton, 
A.~A., Treu, T., et al.\ 2012, \mnras, 422, 3574 

\bibitem[Cervi{\~n}o et al.(2013)]{cervino2013} Cervi{\~n}o, M., Rom{\'a}n-Z{\'u}{\~n}iga, C., Luridiana, V., et al.\ 2013, \aap, 553, A31 

\bibitem[Chandar et al.(2010)]{chandar2010} Chandar, R., Fall, 
S.~M., \& Whitmore, B.~C.\ 2010, \apj, 711, 1263 

\bibitem[Chen et al.(2013)]{chen2013} Chen, X., Arce, H.~G., 
Zhang, Q., et al.\ 2013, \apj, 768, 110 

\bibitem[Chernoff 
\& Shapiro(1987)]{chernoff1987} Chernoff, D.~F., \& Shapiro, S.~L.\ 1987, \apj, 322, 113 

\bibitem[Chernoff \& Weinberg(1990)]{chernoff1990} Chernoff, D.~F., \&
Weinberg, M.~D.\ 1990, \apj, 351, 121

\bibitem[Connelley et al.(2008)]{connelley2008} Connelley, M.~S., 
Reipurth, B., \& Tokunaga, A.~T.\ 2008, \aj, 135, 2526 

\bibitem[Crowther et al.(2010)]{crowther2010} Crowther, P.~A., 
Schnurr, O., Hirschi, R., et al.\ 2010, \mnras, 408, 731 

\bibitem[Dabringhausen et al.(2012)]{dabringhausen2012} Dabringhausen, 
J., Kroupa, P., Pflamm-Altenburg, J., \& Mieske, S.\ 2012, \apj, 747, 72 


\bibitem[de Grijs \& Anders(2012)]{degrijs2012} de Grijs, R., \& Anders, P.\ 2012, \apjl, 758, L22 

\bibitem[de Grijs \& Parmentier(2007)]{degrijs2007}de Grijs, R., \&
Parmentier, G.\ 2007, ChJAA, 7, 155

\bibitem[Duch{\^e}ne \& Kraus(2013)]{duchene2013} Duch{\^e}ne, G., \& Kraus, A.\ 2013, \araa, 51, 269 


\bibitem[Dutton et al.(2012)]{dutton2012} Dutton, A.~A., Mendel, 
J.~T., \& Simard, L.\ 2012, \mnras, 422, L33 

\bibitem[Eggleton et al.(1989)]{eggleton1989} Eggleton, P.~P., Tout,
C.~A., \& Fitchett, M.~J.\ 1989, \apj, 347, 998

\bibitem[Eggleton et al.(1990)]{eggleton1990} Eggleton, P.~P., 
Fitchett, M.~J., \& Tout, C.~A.\ 1990, \apj, 354, 387 


\bibitem[Ferreras et al.(2013)]{ferreras2013} Ferreras, I., La 
Barbera, F., de la Rosa, I.~G., et al.\ 2013, \mnras, 429, L15 


\bibitem[Fukushige 
\& Heggie(2000)]{fukushige2000} Fukushige, T., \& Heggie, D.~C.\ 2000, \mnras, 318, 753 

\bibitem[Gualandris et al.(2004)]{gualandris2004} Gualandris, A., 
Portegies Zwart, S., \& Eggleton, P.~P.\ 2004, \mnras, 350, 615 

\bibitem[Geha et al.(2013)]{geha2013} Geha, M., Brown, T.~M., 
Tumlinson, J., et al.\ 2013, \apj, 771, 29 


\bibitem[Gieles 
\& Baumgardt(2008)]{gieles2008} Gieles, M., \& Baumgardt, H.\ 2008, \mnras, 389, L28 

\bibitem[Gieles \& Portegies Zwart(2011)]{gieles2011} Gieles, M., \& Portegies Zwart, S.~F.\ 2011, \mnras, 410, L6 

\bibitem[Goodwin(1997)]{goodwin1997} Goodwin, S.~P.\ 1997, \mnras, 
286, 669 

\bibitem[Goudfrooij 
\& Kruijssen(2013)]{goudfrooij2013} Goudfrooij, P., \& Kruijssen, J.~M.~D.\ 2013, \apj, 762, 107 

\bibitem[Greissl(2010)]{greissl2010} Greissl, J.~J.\ 2010, 
Ph.D.~Thesis,  

\bibitem[Heggie \& Hut(2003)]{heggiehut} Heggie, D., \& Hut, P.\ 2003,
The Gravitational Million-Body Problem: A Multidisciplinary Approach
to Star Cluster Dynamics. Cambridge University Press


\bibitem[Hurley et al.(2000)]{hurley2000} Hurley, J.~R., Pols, 
O.~R., \& Tout, C.~A.\ 2000, \mnras, 315, 543 

\bibitem[Karl et al.(2011)]{karl2011} Karl, S.~J., Fall, S.~M., 
\& Naab, T.\ 2011, \apj, 734, 11 

\bibitem[Kim et al.(2006)]{kim2006} Kim, S.~S., Figer, D.~F., 
Kudritzki, R.~P., \& Najarro, F.\ 2006, \apjl, 653, L113 


\bibitem[Kouwenhoven et al.(2005)]{kouwenhoven2005} Kouwenhoven,
M.~B.~N., Brown, A.~G.~A., Zinnecker, H., Kaper, L., \& Portegies
Zwart, S.~F.\ 2005, \aap, 430, 137

\bibitem[Kouwenhoven et al.(2007)]{kouwenhoven2007} Kouwenhoven,
M.~B.~N., Brown, A.~G.~A., Portegies Zwart, S.~F., \& Kaper, L.\ 2007,
\aap, 474, 77

\bibitem[Kroupa(2001)]{kroupa2001} Kroupa, P.\ 2001, \mnras, 322, 
231 


\bibitem[Lada \& Lada(2003)]{lada2003} Lada, C.~J., \& Lada, E.~A.\
2003, \araa, 41, 57


\bibitem[Lamers \& Gieles(2006)]{lamers2006letter} Lamers, H.~J.~G.~L.~M., \& Gieles, M.\ 2006, \aap, 455, L17 

\bibitem[Lamers et al.(2005)]{lamers2005} Lamers, H.~J.~G.~L.~M., Gieles, M., \& Portegies Zwart, S.~F.\ 2005, \aap, 429, 173 

\bibitem[L{\"a}sker et al.(2013)]{lasker2013} L{\"a}sker, R., van 
den Bosch, R.~C.~E., van de Ven, G., et al.\ 2013, \mnras, L122 


\bibitem[Maschberger \& Clarke(2008)]{maschberger2008} Maschberger,
T., \& Clarke, C.~J.\ 2008, \mnras, 391, 711


\bibitem[McCrady et al.(2003)]{mccrady2003} McCrady, N., Gilbert, 
A.~M., \& Graham, J.~R.\ 2003, \apj, 596, 240 

\bibitem[McCrady et al.(2005)]{mccrady2005} McCrady, N., Graham,
J.~R., \& Vacca, W.~D.\ 2005, \apj, 621, 278


\bibitem[Mengel et 
al.(2002)]{mengel2002} Mengel, S., Lehnert, M.~D., Thatte, N., \& Genzel, R.\ 2002, \aap, 383, 137 

\bibitem[Mengel et 
al.(2008)]{mengel2008} Mengel, S., Lehnert, M.~D., Thatte, N.~A., et al.\ 2008, \aap, 489, 1091 

\bibitem[Meylan 
\& Heggie(1997)]{meylan1997} Meylan, G., \& Heggie, D.~C.\ 1997, \aapr, 8, 1 

\bibitem[Oey(2011)]{oey2011} Oey, M.~S.\ 2011, \apjl, 739, L46 

\bibitem[Parker \& Goodwin(2007)]{parker2007} Parker, R.~J., \&
Goodwin, S.~P.\ 2007, \mnras, 380, 1271

\bibitem[Plummer(1911)]{plummer1911} Plummer, H.~C.\ 1911, \mnras, 71,
460

\bibitem[Portegies Zwart et  al.(1998)]{portegieszwart1998} Portegies Zwart, S.~F., Hut, P., Makino, J., \& McMillan, S.~L.~W.\ 1998, \aap, 337, 363 

\bibitem[Portegies Zwart et al.(2001)]{portegies2001} Portegies Zwart, S.~F., McMillan, S.~L.~W., Hut, P., \& Makino, J.\ 2001, \mnras, 321, 199 


\bibitem[Portegies Zwart et al.(2010)]{portegieszwart2010} Portegies Zwart, S.~F., McMillan, S.~L.~W., \& Gieles, M.\ 2010, \araa, 48, 431 

\bibitem[Raghavan et al.(2010)]{raghavan2010} Raghavan, D., 
McAlister, H.~A., Henry, T.~J., et al.\ 2010, \apjs, 190, 1 

\bibitem[Reddish(1978)]{reddish1978} Reddish, V.~C.\ 1978, 
International Series in Natural Philosophy, Oxford: Pergamon, 1978

\bibitem[Reipurth et al.(2014)]{reipurth2014} Reipurth, B., Clarke, 
C.~J., Boss, A.~P., et al.\ 2014, arXiv:1403.1907 

\bibitem[Ross et al.(1997)]{ross1997} Ross, D.~J., Mennim, A., \& Heggie, D.~C.\ 1997, \mnras, 284, 811 

\bibitem[Salpeter(1955)]{salpeter1955} Salpeter, E.~E.\ 1955, \apj,
121, 161

\bibitem[Schilbach et 
al.(2006)]{schilbach2006} Schilbach, E., Kharchenko, N.~V., Piskunov, A.~E., R{\"o}ser, S., \& Scholz, R.-D.\ 2006, \aap, 456, 523 


\bibitem[Shin et al.(2013)]{shin2013} Shin, J., Kim, S.~S., 
Yoon, S.-J., \& Kim, J.\ 2013, \apj, 762, 135 

\bibitem[Smith \& Gallagher(2001)]{smith2001} Smith, L.~J., \&
Gallagher, J.~S.\ 2001, \mnras, 326, 1027

\bibitem[Smith 
\& Lucey(2013)]{smith2013} Smith, R.~J., \& Lucey, J.~R.\ 2013, \mnras, 434, 1964 

\bibitem[Spiniello et al.(2012)]{spiniello2012} Spiniello, C., 
Trager, S.~C., Koopmans, L.~V.~E., \& Chen, Y.~P.\ 2012, \apjl, 753, L32 


\bibitem[Sternberg(1998)]{sternberg1998} Sternberg, A.\ 1998, \apj, 
506, 721 

\bibitem[Stolte et al.(2005)]{stolte2005} Stolte, A., Brandner, 
W., Grebel, E.~K., Lenzen, R., \& Lagrange, A.-M.\ 2005, \apjl, 628, L113 

\bibitem[Terlevich(1987)]{terlevich1987} Terlevich, E.\ 1987, \mnras, 224, 193 

\bibitem[Tout et al.(1997)]{tout1997} Tout, C.~A., Aarseth, 
S.~J., Pols, O.~R., \& Eggleton, P.~P.\ 1997, \mnras, 291, 732 

\bibitem[Vanbeveren(1982)]{vanbeveren1982} Vanbeveren, D.\ 1982, \aap,
115, 65


\bibitem[Weidner \& Kroupa(2006)]{weidnerkroupa2006} Weidner, C., \&
Kroupa, P.\ 2006, \mnras, 365, 1333

\bibitem[Weidner et al.(2013)]{weidner2013} Weidner, C., Kroupa, 
P., Pflamm-Altenburg, J., \& Vazdekis, A.\ 2013, \mnras, 2503 

\bibitem[Zaritsky et al.(2012)]{zaritsky2012} Zaritsky, D., Colucci, 
J.~E., Pessev, P.~M., Bernstein, R.~A., \& Chandar, R.\ 2012, \apj, 761, 93 


\end{thebibliography}
\end{document}